\documentclass{statsoc}
\usepackage[a4paper]{geometry}
\usepackage{graphicx}
\usepackage[textwidth=8em,textsize=small]{todonotes}
\usepackage{pdfsync}
\usepackage{natbib}
\usepackage{amsmath,amssymb}  
\usepackage{dsfont}
\usepackage{bm}
\graphicspath{{./figures/}}
\usepackage{url}

\DeclareMathOperator{\J}{\mathcal{J}}
\def\V{\mathcal{V}}
\def\E{\mathcal{E}}
\def\G{\mathcal{G}}
\newcommand{\e}{\mathds{E}}
\newcommand{\var}{\mathds{V}\text{ar}}
\newcommand{\cov}{\mathds{C}\text{ov}}
\newcommand{\bias}{\mathds{B}}

\DeclareMathOperator{\T}{^{\mathsf T}}
\DeclareMathOperator{\w}{\boldsymbol{\omega}}
\def\V{\mathcal{V}}
\def\E{\mathcal{E}}
\def\G{\mathcal{G}}

\newcommand{\indep}{\mathrel{{\perp}\hspace*{-0.6em}{\perp}}}
\newcommand{\given}{\mathrel{|}}
\newcommand{\condindep}[3]{#1 \indep #2 \given #3}
\title[Multivariate spatial hybrid processes]{A spatial dependence graph model for multivariate spatial hybrid processes}
\author[M. Eckardt and J. Mateu]{Matthias Eckardt}
\address{Department of Mathematics, 
Universitat Jaume I,
 Castell{\'o}n, Spain.}
\email{eckardtm@cms.hu-berlin.de}
\author[M. Eckardt and J. Mateu]{Jorge Mateu}
\address{Department of Mathematics, 
Universitat Jaume I,
 Castell{\'o}n, Spain.}
 
\begin{document}
\maketitle
\begin{abstract}
This paper is concerned with the joint analysis of multivariate mixed-type spatial data, where some components are point processes and some are of lattice-type by nature. After a survey of statistical methods for marked spatial point and lattice processes, the class of multivariate spatial hybrid processes is defined and embedded within the framework of spatial dependence graph models. In this model, the point and lattice sub-processes are identified with nodes of a graph whereas missing edges represent conditional independence among the components. This finally leads to a general framework for any type of spatial data in a multivariate setting.  We demonstrate the application of our method in the analysis of a multivariate point-lattice pattern on crime and ambulance service call-out incidents recorded in London, where the points are the locations of different pre-classified crime events and the lattice components report different aggregated incident rates at ward level.

\end{abstract}
%%%%%%%%%%%%%%%%%%%%%%%%%%%%%%%%%%%%%%%%%%%%%%%%%%%%%%%%%%%%%%%%%%%%%%%%%%%%%%%%%%%%%%%%%%%%%%%%%%%
\section{Introduction}

Stimulated by the enormous technological and scientific progress, the statistical analysis of spatial data is a rapidly developing field which concerns the exploration and characterisation of potential structures and interrelations among a set of observations recorded in some bounded planar observation window. While various criminological, ecological, epidemiological or environmental research questions have been addressed, the heterogeneity of scientific perspectives has led to a great variety of spatial data specifications and statistical techniques for (a) point-referenced,  (b) spatial lattice and (c) marked spatial point patterns. 

The expeditious increase of information technologies and storage capacities has led to a plethora of multivariate data on numerous outcomes in space. Although a considerable body of literature exists on spatial data and spatial data analysis, and several authors have contributed to this field, the need for efficient techniques which jointly detect the global conditional structural interrelations in a multivariate spatial data setting  still remains. The growing availability and accessibility of multivariate spatial data as well as the rapid developments in geographical information systems (GIS) have led to an ever-increasing demand for statistical methods and computationally efficient tools that not only account for the inherent complexity and structural interrelations of such data but also facilitate a clear interpretation. Although a limited number of methodological contributions on multivariate spatial interrelations exists, including the work of \cite{Diggle2005}, \cite{Cressie2015}, \cite{Genton2015}, \cite{Grabarnik2009}   \cite{Guinness2014}, \cite{Illian2007}, \cite{Shimatani2001} and \cite{Waagepetersen2016}, this demand for efficient statistical techniques has hardly been satisfied. Specifically, there is an emerging need for efficient exploratory tools which allow for the simultaneous analysis of conditional cross-type interrelations among different components in multivariate spatial data. Notable exceptions have just recently been proposed by \cite{Eckardt2016} by means of the spatial dependence graph model for qualitatively marked (commonly called multivariate or multi-type) patterns and extended to the case where both,  qualitative and quantitative marks are available (a multivariate‐marked spatial point process) by \cite{EckardtMateuSDGM}.

While almost all statistical treatments of geostatistical, lattice-type and point patterns have run in parallel and each type of data has been investigated separately, one might be interested in exploring potential interrelations between different types of spatial data, e.g. between different point- and lattice-type components in a multivariate setting. However, although some authors have contributed to the joint analysis of time series and point process in the temporal domain including  \cite{brillinger1994}, \cite{HALLIDAY1995}, \cite{Henschel2008} and \cite{rigas1983}, mixtures of spatial point processes and spatial lattice data (so-called spatial hybrids) have not been studied much so far. Exceptions such as \cite{Augustin}, \cite{Kanaan2000}, and \cite{Kanaan2008} remain restricted to at most the bivariate case considering mixtures of one unmarked point and one lattice component. Inspired by the work on spatial graphical models for multitype and multivariate-marked point processes of \cite{Eckardt2016} and \cite{EckardtMateuSDGM} and some developments in the analysis of irregularly-spaced time series presented by \cite{Bauwens2009}, \cite{Engel1998} and \cite{Hasbrouck1991}, this paper aims to contribute to the multivariate analysis of spatial data. In particular, a unifying approach based on partial marked point characteristics is developed which allows for the simultaneous analysis of any type of multivariate spatial data by means of an undirected graphical model.

This paper is structured as follows. Section \ref{sec:1} presents the basic properties of spatial point and lattice processes in the spatial and frequency domain. The class of spatial hybrid processes is discussed and extended to multivariate mixed-type processes in Section 3 yielding the definition of a spatial dependence graph model for hybrid data. An application of the proposed model to mixed-type data on crime events and aggregated ambulance service call-outs is given in Section \ref{sec:2}.  Finally, the paper ends with some conclusions and a discussion. 

\section{Recapitulating spatial point and lattice process characteristics}\label{sec:1}

To introduce a general framework for multivariate spatial hybrid processes, the fundamental properties of point and lattice-type processes need to be recapitulated first. 

\subsection{Recapitulating spatial point process characteristics}

This section presents a short summary of first and second-order properties of spatial point processes in the spatial domain. For an in-depth treatment of the subject, we refer the interested reader to \cite{Chiu2013}, \citet{Diggle2002},  \cite{Illian2008}, \cite{Moller2004} and \cite{StoyanStoyan1994}.  

\subsubsection{First- and second-order characteristics in the spatial domain}
Usually, the first-order properties of a spatial point process are expressed by means of the first-order intensity function. Adopting the notation of \citet{Diggle2002, Diggle2013}, the first- and second-order intensity functions are given as 
\[
\lambda_i(\mathbf{s})=\lim_{|d\mathbf{s}|\rightarrow 0}\left\{\frac{\e\left[ N_i(d\mathbf{s}))\right]}{|d\mathbf{s}|}\right\}, \mathbf{s}\in\mathbf{S}.
\]  
and 
\[
\lambda_{ii}(\mathbf{s,s'})=\lim_{|d\mathbf{s}|,|d\mathbf{s}|\rightarrow 0}\left\{\frac{\e\left[N_i(d\mathbf{s})N_i(d\mathbf{s'})\right]}{|d\mathbf{s}||d\mathbf{s'|}}\right\}, \mathbf{s}\neq\mathbf{s'}, \mathbf{s},\mathbf{s'}\in\mathbf{S},
\]
respectively. Here, $\mathbf{s}=(x, y)$ and $\mathbf{s'}=(x',y')$ are the location of two distinct randomly occurring events within a bounded region $\mathbf{S}\subset\mathds{R}^2$, $N_i(d\mathbf{s})$ and $N_i(d\mathbf{s}')$ with $ N_i(d\mathbf{s}) = N_i(\mathbf{s}+d\mathbf{s}) - N_i(\mathbf{s})$ are the number of observed events of type $i$ and type $j$ within two infinitesimal discs containing $\mathbf{s}$ and $\mathbf{s}'$, respectively, and $|\cdot|$  denotes the area of the argument. Apart from the second-order intensity function, which is closely connected to Ripleys' $K$-function \citep{Ripley1976}, another important characteristic is the covariance density function $\gamma(\mathbf{s})$. In particular, in the multivariate setting where different types of points are observed within a congruent window, two versions of $\gamma(\mathbf{s})$ are of interest: (a) the auto- and (b) the cross-covariance density function defined by
\begin{equation}
\gamma_{ii}(\mathbf{s,s'})=\lim_{|d\mathbf{s}|,|d\mathbf{s'}|\rightarrow 0}\left\{\frac{\e\left[\lbrace N_i(d\mathbf{s})-\lambda_i(d\mathbf{s})\rbrace \lbrace N_i(d\mathbf{s'})-\lambda_i(d\mathbf{s'})\rbrace\right]}{|d\mathbf{s}||d\mathbf{s'|}}\right\}
\label{autocov}
\end{equation}
and 
\begin{equation}
\gamma_{ij}(\mathbf{s,s'})=\lim_{|d\mathbf{s}|,|d\mathbf{s'}|\rightarrow 0}\left\{\frac{\e\left[\lbrace N_i(d\mathbf{s})-\lambda_i(d\mathbf{s})\rbrace\lbrace N_j(d\mathbf{s'})-\lambda_j(d\mathbf{s'})\rbrace\right]}{|d\mathbf{s}||d\mathbf{s'|}}\right\},
\label{crosscov}
\end{equation}
respectively. 

However, under orderliness, we have $\e\left[\lbrace N_i(d\mathbf{s})\rbrace^2\right]=\lambda_i(\mathbf{s})|d\mathbf{s}|$ whenever $\mathbf{s}=\mathbf{s'}$. This problem is solved by  including this expression into \eqref{autocov} yielding  Bartletts' complete (auto)-covariance density function $\kappa_{ii}(\cdot)$  \citep{Bartlett1964}, namely
\begin{equation}\label{completecov}
\kappa_{ii}(\mathbf{s,s'})=\lambda_i(\mathbf{s})\delta(\mathbf{s}-\mathbf{s'})+\gamma_{ii}(\mathbf{s,s'})
\end{equation}
where $\delta(\cdot)$ denotes a two-dimensional Dirac delta function. Again, focussing on a multivariate setting, both the complete auto- and the complete cross-covariance density functions could be defined where, adopting the result of   \cite{Mugglestone1996a}, we set $\kappa_{ij}(\mathbf{s,s'})=\gamma_{ij}(\mathbf{s,s'})$ and $\kappa_{ji}(\mathbf{s,s'})=\gamma_{ji}(\mathbf{s,s'})$.

While the above characteristics are defined with respect to multivariate spatial point patterns, e.g. when different types of points are available, the mean product of marks $U(r)$ for points separated by the distance $r$, which is an important characteristic for the case when additional integer-valued are available for each type of points, is treated next.
In the univariate case, adopting the notation of \cite{Capobianco1998}, $U(r)$ is defined by
\begin{equation}\label{eq:meanmarks}
U(r)=\lambda^2g(r)k_{mm}(r)d\mathbf{s}d\mathbf{s}'
\end{equation}
where $g(r)$ and $k_{mm}(r)$ are the pair and mark correlation functions, respectively, as described e.g. by \cite{Illian2008}. For a discussion of alternative formulation of the mean product of marks we refer the interested reader to \cite{Capobianco1998} and \cite{EckardtMateuSDGM}. 

\subsubsection{Spectral properties of multivariate spatial point processes}
\label{sec:sppSpectraTheo}

Next, point process characteristics defined in the frequency domain are discussed. These frequency domain characteristics are based on Fourier transformations of (marked) point locations to matrices of (marked) auto- and cross-periodogram values and spectral analysis techniques. The elements of the estimated auto- and cross-spectra matrices, the so-called ordinates, hold information about the strength of periodicities in the auto- and cross-covariance density functions of the underlying point process.  For simplicity of the spectral expressions, we only discuss the second-order stationary case.  We remark that although spectral techniques have become a prominent tool for the analysis of time series data and certain advantages exist, these techniques have not been studied and applied to spatial point processes much so far and the number of methodological and applied contributions remain limited.
 
The content presented here can be understood as a straightforward extension of the spectral analysis of point events in the temporal domain, as described by \citet{Bartlett1963} and \citet{Brillinger1972}, to the two-dimensional spatial case first presented by \citet{Bartlett1964}. Further contributions to the spectral analysis of spatial point processes can be found in the papers of  \cite{Mugglestone1990}, \cite{Mugglestone1996b, Mugglestone1996a,Mugglestone2001},  \cite{Renshaw1997,Renshaw2002}, \cite{Renshaw1983,Renshaw1984} and \cite{saura2006}  which serve as   fundamental references in this section.

For a second-order stationary multivariate spatial point process, the auto-spectral density function (the auto-spectrum) for points of type $i$ at frequencies $\mathbf{w}=( w_1, w_2)$ appears as the Fourier transform of the complete auto-covariance density function $\kappa_{ii}$ of $N_i$,  
 \begin{equation}\label{fouriereq}
\begin{split}
f_{ii}(\mathbf{w}) &= \int \kappa_{ii}(\mathbf{c})\exp(-\imath\mathbf{w}^{\T}\mathbf{c})d\mathbf{c}\\
&=\int^\infty_{-\infty}\int^\infty_{-\infty}\kappa_{ii}(c_1,c_2)\exp\lbrace-\imath( w_1c_1+ w_2c_2)\rbrace dc_1dc_2
\end{split}
\end{equation}
where $\imath=\sqrt{-1}$, $\mathbf{c}=(c_1,c_2)$ with $c_1=x-x^\prime$ and $c_2=y-y^\prime$, and $\mathbf{w}^{\T}$ denotes the transpose of $\mathbf{w}$. 
From \eqref{fouriereq}, the complete auto-covariance density function can uniquely be recovered via the inverse Fourier transformation of $f_{ii}(\mathbf{w})$,
\begin{equation}\label{inversekappa}
\kappa_{ii}(\mathbf{c}) = \int f_{ii}(\mathbf{w})
\exp\left(\imath\mathbf{w}^{\T}\mathbf{c}\right)d\mathbf{w}.
\end{equation}
As described in \citet{Brillinger1981} and \cite{Brockwell2006} with respect to time series, the auto-spectrum can be understood as the decomposition of $\kappa_{ii}$ into a periodic function of frequencies $\mathbf{w}$.
Substituting \eqref{completecov} into \eqref{fouriereq} leads to 
 \begin{equation}\label{fouriereqfinal}
f_{ii}(\mathbf{w}) =\lambda_i+\int^\infty_{-\infty}\int^\infty_{-\infty}\zeta_{ii}(c_1,c_2)\exp\lbrace-\imath( w_1c_1+ w_2c_2)\rbrace dc_1dc_2.
\end{equation}

Likewise, the cross-spectral density function (the cross-spectrum) is obtained as the Fourier transform of the complete cross-covariance density function $\kappa_{ij}$,
\begin{equation}\label{crossspectrakappa}
\begin{aligned}
f_{ij}(\mathbf{w}) &= \int \kappa_{ij}(\mathbf{c})\exp(-\imath\mathbf{w}^{\T}\mathbf{c})d\mathbf{c}\\
 &= \int \zeta_{ij}(\mathbf{c})\exp(-\imath\mathbf{w}^{\T}\mathbf{c})d\mathbf{c},
\end{aligned}
\end{equation}
and measures the linear interrelation of components $N_i$ and $N_j$. Two processes are said to be uncorrelated at all spatial lags if and only if the corresponding spectrum is zero at all frequencies. Recalling that $\kappa_{ij}(\mathbf{c})=\kappa_{ji}(-\mathbf{c})$, we also have $f_{ij}(\mathbf{w})=f_{ji}(-\mathbf{w})$ from which we deduce that it suffices to consider only one cross-spectrum (cf. \cite{Bartlett1964} and \cite{Mugglestone1996b, Mugglestone1996a}). 

Notice that as $\zeta_{ij}(\mathbf{c})\neq \zeta_{ij}(\mathbf{-c})$, the cross-spectrum is a complex-valued function and thus a decomposition  of the complex-valued cross-spectrum into the real and the imaginary parts in terms of the co-spectrum $C_{ij}(\mathbf{w})$ and the quadrature spectrum $Q_{ij}(\mathbf{w})$ using a Cartesian coordinates representation or in terms of its modulus $\mathfrak{a}_{ij}(\mathbf{w})$ and phase $\wp_{ij}(\mathbf{w})$ using a decomposition into polar coordinates is applied. While $\mathfrak{a}_{ij}(\mathbf{w})$ measures the relative magnitude of the power attributable to  frequencies $\mathbf{w}$ in a bivariate point pattern, $\wp_{ij}(\mathbf{w})$ indicates how closely linear translations of the pattern formed by one component match the pattern formed by the other component. In this respect, the cross-phase spectrum  measures the similarity of two patterns up to linear shifts (cf. \citet{ Chatfield1989, Priestley1981}). This information is provided by the slope of the cross-phase which measures the magnitude and direction of the shift. Obviously,  $\wp_{ij}(\mathbf{w})$ is undefined whenever the cross-spectrum vanishes and its meaning is questionable if only small values of the cross-spectrum appear.

In the motion-invariant case such that the process is invariant under rotation and translation, \cite{Bartlett1964} showed a simplification of both spectral expression based on a polar coordinates representation of \eqref{fouriereqfinal} and \eqref{crossspectrakappa} yielding   
\begin{equation}
\label{eq:isotropicautospectra}
f_{ii}(\varpi)=\lambda_i+ 2\pi \int_0^\infty r\zeta_{ii}(r)\J_0(r\varpi)dr
\end{equation}
and 
\begin{equation}
\label{eq:isotropiccrossspectra}
\begin{aligned}
f_{ij}(\varpi)&=2\pi \int_0^\infty r\zeta_{ij}(r)\J_0(r\varpi)dr \\ 
&= C_{ij}(\varpi)
\end{aligned}
\end{equation}
where $\J_0(r\varpi)=(2\pi)^{-1}\int_{-\pi}^\pi\exp(r\varpi \sin u)$  is the unmodified Bessel function of the first kind of order zero as described in \citet{Watson1944} and $\varpi=\sqrt{ w_1^2+ w_2^2}$. 
For the cross-spectral term, we note that from  \eqref{eq:isotropiccrossspectra} it follows that the phase spectrum $\wp_{ij}(\varpi)$ and the quadrature spectrum $Q_{ij}(\varpi)$ are identically zero at all frequencies. Further, for the cross-amplitude  spectrum $\mathfrak{a}_{ij}(\varpi)$ we have  
\begin{equation}
\begin{aligned}
\mathfrak{a}_{ij}(\varpi)&= \mod(f_{ij}(\varpi))\\
&=2\pi \left|\int_0^\infty r\zeta_{ij}(r)\J_0(r\varpi)dr\right|
\end{aligned}
\end{equation}
(cf. \cite{Mugglestone1996b, Mugglestone1996a}).

Before we discuss the estimation of the introduced terms from observed point locations, we first concern the spectral coherence $\vert R_{ij}(\mathbf{w})\vert^2$ which is  defined as a rescaled version of the cross-spectrum, 
\begin{equation}\label{eq:Coh}
\vert R_{ij}(\mathbf{w})\vert^2=\frac{f_{ij}(\mathbf{w})^2}{\left[f_{ii}(\mathbf{w})f_{jj}(\mathbf{w})\right]},
\end{equation}
and provides a measure on the linear relation between two components. Different from the auto- and cross-spectra we have that $0\leq \vert R_{ij}(\mathbf{w})\vert^2\leq 1$. We note that the quantity $R_{ij}(\mathbf{w})$ whose modulus squared is the spectral coherence is called the spectral coherency (cf. \cite{Priestley1981}).

\subsubsection{Estimation of spectral density functions for multivariate point patterns}\label{sec:DFTspp}

Whilst the previous Section briefly revises the formal definitions of the auto- and cross-spectral density functions, the estimation of both functions from a $d$-variate spatial point pattern is considered next where we assume that $\mathbf{n}=(n_1,\ldots,n_d)$ points are observed within a rectangular region $\mathbf{S}\subset\mathds{R}^2$ with sides of lengths $l_1$ and $l_2$. Writing $\lbrace\mathbf{s}_i\rbrace=\lbrace(x_i,y_i)\rbrace, i=1,\ldots,n_i$ for the locations of points for component $i$ and $\lbrace\mathbf{s}_j\rbrace$ for the locations of points for component $j$, both empirical spectra can be obtain through a discrete Fourier transforms (DFT) of the point locations itself where the DFT of  $\lbrace\mathbf{s}_i\rbrace$ is defined as
\begin{equation}\label{eq:DFTspp}
\begin{aligned}
\mathcal{F}_i(p,q)&=\frac{1}{\sqrt{l_1l_2}}\sum_{i=1}^{n_i}\exp\left(- 2\pi\imath n_i^{-1}\left(px_i + qy_i\right)\right)\\
&= a_i(p,q)+\imath b_i(p,q).
\end{aligned}
\end{equation} 
Here, $a_i(p,q)$ and $b_i(p,q)$ are the real and the imaginary parts of $\mathcal{F}(p,q)$. From this expression, the auto-periodogram itself for frequencies $\mathbf{w}=(2\pi p/n_i,2\pi q/n_i)$ is obtained as 
\begin{eqnarray}\label{eq:autoperiodogram}
\widehat{f}_{ii}(\mathbf{w})&=&\mathcal{F}_i(p,q)\overline{\mathcal{F}}_i(p,q)\\
&=& \lbrace a_i(p,q)\rbrace^2+\lbrace b_i(p,q)\rbrace^2\nonumber
\end{eqnarray}
where $\overline{\mathcal{F}}_i$ denotes the complex conjugate of $\mathcal{F}_i$. 

We note that in case of complete spatial randomness, as pointed out by \cite{Mugglestone1990} and \cite{Kanaan2000}, the bias $\bias$ of \eqref{eq:autoperiodogram} is
\[
\bias(\mathbf{w})= 2l_1l_2\lambda_i^2\left[\frac{\sin\left(\frac{l_1 w_p}{2}\right)}{\left(\frac{l_1 w_p}{2}\right)}\times\frac{\sin\left(\frac{l_2 w_q}{2}\right)}{\left(\frac{l_2 w_q}{2}\right)}\right]^2.
\]

To avoid bias at low frequencies, the point locations $\lbrace\mathbf{s}_i\rbrace$ are usually replaced by the standardised coordinates $\lbrace\left(x_i^{\ast}, y_i^{\ast}\right)\rbrace,~i=1,\ldots,n_i$ with $x_i^{\ast}=n_ix_i/l_1$ and $y_i^{\ast}=n_iy_i/l_2$ prior to the analysis (cf. \citet{Bartlett1964, Mugglestone1996a}). If the periodograms are computed from unstandardised coordinates, their values will start to repeat after $n_i$ rows and/ or columns. This phenomenon is commonly called aliasing. In case of aliasing, it becomes impossible to decide whether or not high frequencies are present in the spatial point pattern. 
An equivalent technique to avoid bias at low frequencies is to rescale the point pattern to the unit square. In this particular case,  \eqref{eq:DFTspp} reduces to 
\begin{equation}\label{eq:sppShortform}
\mathcal{F}_i(p,q)=\sum^{n_i}_{i=1}\exp(-2\pi\imath(px_i+qy_i)).
\end{equation}

The cross-periodogram is computed analogous to \eqref{eq:autoperiodogram} by using the following expression
\begin{equation}\label{eq:crossperiodogram}
\widehat{f}_{ij}(\mathbf{w})=\mathcal{F}_i(p,q)\overline{\mathcal{F}}_j(p,q).
\end{equation}
Decomposing the cross-periodogram into the real and the imaginary parts leads to the co- and quadrature spectra
\[
\widehat{C}_{ij}(\mathbf{w})=a_i(p,q)a_j(p,q)+b_i(p,q)b_j(p,q)
\]
and
\[
\widehat{Q}_{ij}(\mathbf{w})=b_i(p,q)a_j(p,q)- a_i(p,q)b_j(p,q).
\]

We note that as two point locations can lie arbitrarily close together, the maximum frequency that can be resolved, the so-called Nyquist frequency, is infinite and any range of scales of the point pattern might be considered. Recalling that $f_{ij}(\mathbf{w})=f_{ji}(-\mathbf{w})$ and $f_{ii}(\mathbf{w})=f_{ii}(-\mathbf{w})$, the corresponding periodograms $\widehat{f}_{ii}(\mathbf{w})$ and $\widehat{f}_{ij}(\mathbf{w})$ are also symmetric and it suffices to compute the periodograms over both negative and positive integers for one of the frequency coordinates, say $q$, and consider only positive integers for the alternative coordinate, $p$. The information on the frequencies related to the negative values of $p$ are then obtained using the symmetry property of the periodograms. As the choices for $p$ and $q$ are left to the user, \cite{Renshaw1983} and \cite{Mugglestone1996a} suggested to consider the ranges $p = 0, 1, \ldots, 16$ and $q = - 16, \ldots, 15$ for the computation of the auto- and cross-periodogram which are said to provide an adequate cover of frequencies at which structure may be present in the peridogram. In this case, the maximum frequency amplitude of the periodogram is  $\mathbf{w}_{max} = \sqrt{(32\pi/l_1)^2 + (32\pi/l_2)^2}$. If the periodogram is computed from rescaled coordinates, we have $\mathbf{w}_{max} \simeq 23\times 2\pi$ (cf. \cite{Renshaw1983}, \cite{Mugglestone1996a}).

We remark that the auto- and cross-periodograms are  asymptotically unbiased but inconsistent estimates and smoothing is required (cf. \cite{Mugglestone1996b}). 

Estimates for \eqref{eq:Coh} were then obtained by replacing $f_{ii}, f_{jj}$ and $f_{ij}$ by their smoothed empirical counterparts $\widehat{f}_{ii}, \widehat{f}_{jj}$ and $\widehat{f}_{ij}$. As pointed out by \cite{Priestley1981}, we note that if the spectral coherence is computed from raw auto- and cross-periodograms, $|R_{ij}(\mathbf{w})|^2$ is unity at all frequencies as we are effectively computing a correlation coefficient from a single pair of observations at each frequency.

\subsubsection{Spectral properties of multivariate-marked spatial point patterns}\label{sec:spectralpropertiesMMSPP}

The auto- and cross-spectral density functions for MMSPP are defined analogous to the auto- and cross-spectral density functions of multivariate spatial point patterns as treated in Section \ref{sec:sppSpectraTheo}. Extending the results of \cite{Renshaw2002} to the multivariate case, the marked auto- and cross-spectral density functions are obtained by replacing the complete auto- and cross-covariance density functions of \eqref{fouriereq} and \eqref{crossspectrakappa} by the auto- and cross-type mean product of marks.
Then, the marked auto-spectral density function follows as 
\begin{equation}
\label{eq:fourierautoMMSPP}
f_{ii}(\mathbf{w}) = \int U_{ii}(\cdot)\exp(-\imath\mathbf{w}\T\mathbf{c})d\mathbf{c}.
\end{equation}
Similarly, for the marked cross-spectral density function we have
\begin{equation}
\label{eq:fouriercrossMMSPP}
f_{ij}(\mathbf{w}) = \int U_{ij}(\cdot)\exp(-\imath\mathbf{w}\T\mathbf{c})d\mathbf{c}.
\end{equation}
We note  that the explicit expressions for $U_{ii}(\cdot)$ and $U_{ij}(\cdot)$ depend on the specification of the auto- and cross-type versions of \eqref{eq:meanmarks} as discussed by \cite{EckardtMateuSDGM}.

As in the multivariate case, the cross-type  mean product of marks of \eqref{eq:fouriercrossMMSPP} is not necessarily symmetric and can be decomposed either in terms of Cartesian coordinates or by means of polar coordinates. 

\subsubsection{Estimation of spectral density functions for multivariate point patterns}\label{sec:DFTmspp}

Extending the results of Section \ref{sec:DFTspp} to the multivariate-marked case, the estimation of the empirical marked auto- and cross-spectra is briefly described. As for the multivariate case, both functions can be computed through a DFT  of the marked locations $\lbrace\mathbf{s}_i, m_i(\mathbf{s}_i)\rbrace$ and $\lbrace\mathbf{s}_j, m_j(\mathbf{s}_j)\rbrace$ where 
\begin{equation}\label{ed:DFTmarkNoScaled}
\begin{aligned}
\mathcal{F}_i(p,q)&=\frac{1}{\sqrt{l_1l_2}}\sum_{i=1}^{n_i}\left(m_i(\mathbf{s}_i)-\mu_M(\mathbf{s}_i)\right)\exp\left(- 2\pi\imath n_i^{-1}\left(px_i + qy_i\right)\right)\\
&= a_i(p,q)+\imath b_i(p,q)
\end{aligned}
\end{equation}
is the DFT of the marked locations for points of type $i$.
Here, $m_i(\mathbf{s}_i)$ is the mark for the $i$-th location of component $i,~\mu_M(\mathbf{s}_i))$ is the mean over all marks $m_i(\cdot)$ for locations of type $i$ and $p$ and $q$ are the same as already defined in Section  \ref{sec:DFTspp} (cf. \cite{Renshaw2002}). If the marked locations have been scaled to the unit square, this expression reduces to
\begin{equation}\label{ed:DFTmark}
\mathcal{F}_i(p,q)=\left(\sum_{i=1}^{n_i}\left(m_i(\mathbf{s}_i)-\mu_M(\mathbf{s}_i)\right)\exp(-2\pi\imath(px_i+qy_i))\right).
\end{equation}

\subsection{Recapitulating spatial lattice processes}

We now briefly revise the basic characteristics for spatial lattice processes as described in \citet[Chapter 3]{Banerjee2004}, \citet[Chapter 6]{Cressie1993} and \citet[Chapter 5]{Ripley1981}. In general,  the spatial lattice is assumed to be finite. Let $\lbrace x(\mathbf{s}_n)\rbrace$ denote the realisations of a spatial lattice process $\lbrace X(\mathbf{s}_n)\rbrace$ on $\mathbf{S}_L\subseteq \mathds{Z}^2$. The exposition begins with regularly-shaped spatial lattices where $\mathbf{S}_L$ is assumed to be a rectangular region of dimension $\left[0,l_1\right]\times\left[0,l_2\right]$.  The lengths $l_1$ and $l_2$ are assumed to be integers, however, we do not necessarily require that $l_1=l_2$.  Any measurement made at $X(\mathbf{s})$ on $\mathbf{S}_L$ is associated with a grid square $\left[s_1,s_1+1\right]\times\left[s_2,s_2+1\right]$ recorded along a regular grid of size $s_1=0,\ldots,l_1-1$ and $s_2=0,\ldots,l_2-1$. 

\subsubsection{First- and second-order properties of spatial lattice processes}

Usually, any such process can be characterised by the first-order moment $\e\left[X(\mathbf{s})\right]=\mu(\mathbf{s})$ and the covariance density function $\zeta(\mathbf{s},\mathbf{s}^\prime)=\cov\left[X(\mathbf{s}),X(\mathbf{s}^\prime)\right]$. Under stationarity of  $X(\mathbf{s})$ which implies that the characteristics are invariant under translation, the moments simplify to $\e\left[X(\mathbf{s})\right]=\mu,~\var\left[X(\mathbf{s})\right]=\sigma^2$ and $\cov\left [X(\mathbf{s}),X(\mathbf{s}+\mathbf{c})\right]=\zeta(\mathbf{c})$.  In the remainder, it is assumed that the lattice process is corrected for its mean such that $\zeta(\mathbf{c})=\e\left[X(\mathbf{s})X(\mathbf{s}+\mathbf{c})\right]$ as $\mu=0$. 

Likewise, a multivariate regularly-shaped spatial lattice process $\lbrace\mathbf{X}(\mathbf{s}_n)\rbrace$ is understood as a collection of $d$ disjoint component processes $\lbrace X_i(\mathbf{s}_n)\rbrace, i=1,\ldots,d$, each of which with mean $\mu_i(\mathbf{s})$. The auto- and cross-covariance density functions of $\lbrace X_i(\mathbf{s})\rbrace$ and $\lbrace X_j(\mathbf{s})\rbrace$ are denoted by $\zeta_{ii}(\cdot)=\var\left[X_i(\mathbf{s})\right]$ and $\zeta_{ij}(\cdot)=\cov\left [X_i(\mathbf{s}),X_j(\mathbf{s}^\prime)\right]$, respectively. In the remainder of this section, $\lbrace\mathbf{X}(\mathbf{s})\rbrace$ is assumed to be stationary such that all component processes are marginally and jointly stationary.

Before we turn to the classical analysis of spatial lattice processes, the lattice-type analogue of complete spatial randomness has to be presented first. For the univariate case, a spatial lattice process is said to exhibit complete spatial randomness if all random variables $X(\mathbf{s})$ are \textit{i.i.d.} Gaussian distributed with mean zero and variance $\sigma^2$. Any such process is commonly denoted as Gaussian white noise. For the multivariate case, complete spatial randomness implies that all components $X_i(\mathbf{s}), i= 1,\ldots, d$ are Gaussian white noise. This implies that $\zeta_{ij}(\cdot)=0$ for any two components $i$ and $j$ of $\lbrace\mathbf{X}(\mathbf{s}_n)\rbrace$.

\subsubsection{Spectral properties of spatial lattice processes}

We now consider the characterisation of regularly- and irregularly-shaped spatial lattice processes through the frequency domain using spectral density functions. 

For a stationary multivariate regularly-shaped lattice process, the auto-spectral density function for component $i$ at frequencies $\mathbf{w}=( w_1, w_2)$ is defined as the Fourier transform of the auto-covariance density function  $\zeta_{ii}$,
\[
f_{ii}(\mathbf{w})=\sum_{\mathbf{c}}\zeta_{ii}(\mathbf{c})\exp(-\imath\mathbf{w}\mathbf{c}\T).
\]
The cross-spectral density function (the cross-spectrum) is obtained analogous to the Fourier transform of the cross-covariance density function $\zeta_{ij}$,
\[
f_{ij}(\mathbf{w})=\sum_{\mathbf{c}}\zeta_{ij}(\mathbf{c})\exp(-\imath\mathbf{w}\mathbf{c}\T).
\]
Since $\zeta_{ij}(\mathbf{c})=\zeta_{ji}(-\mathbf{c})$ under stationarity of $\mathbf{X}(\mathbf{s})$, we have $f_{ij}(\mathbf{w})=f_{ji}(-\mathbf{w})$. We note that the cross-spectrum is a complex-valued function and a common procedure is to decompose the complex-valued spectrum using Cartesian or polar coordinates.

As in  the classical analysis of time series, the fraction $f_{ii}(\mathbf{w})/\sigma^2$ can be understood as the mean proportion of the total power of the components with frequencies between $\mathbf{w}$ and $\mathbf{w}+d\mathbf{w}$ (see \cite{Priestley1981}). Notice that, as  $f_{ii}(\mathbf{w})=\sigma^2$  under Gaussian white noise, the theoretical auto-spectral density function equals $1$ under CSR. 

Although regularly-shaped spatial lattice processes are most closely related to time series, they are less important for practical applications where observations are most commonly associated with polygon entities. This type of spatial processes, irregularly-shaped spatial lattice processes, will be covered next.

To start, consider a set of $n$ irregularly-shaped sites, e.g. a set of $n$ polygon entities. Different from regularly-shaped lattice processes, let $x(\mathbf{s}_i)$ denote the measurement made at the centroid $\mathbf{s}_i=(x_i,y_i)$ of the $i$-th irregularly-shaped site. By analogy with the analysis of irregularly-spaced time series, any such sequence can be analysed using classic spatial tools for marked point processes. To this end, the observation $x(\mathbf{s}_i)$ is considered as a quantitative mark $m(\mathbf{s}_i)$ of centroid $\mathbf{s}_i$.  We note that this linkage to  (multivariate-) marked spatial point processes also holds for multivariate regularly-shaped spatial lattice processes if the centroids of regular grid squares are treated as point locations. 

\subsubsection{Estimation of spectral densities for multivariate lattice patterns}

We now concern the estimation of the auto- and cross-spectral density functions from regularly- and irregularly-shaped lattice patterns where regularly-shaped patterns are considered first. Different from the non-parametric estimation presented here, both sample spectra could also be computed through a parametric Whittle approximation \citep{Whittle1954} of a Gaussian log-likelihood  as implemented in the papers of \cite{Guinness2014} and \cite{Terres2015} for regularly-shaped spatial lattice data on soil concentration. 

Suppose we observed a  $d$-variate spatial lattice pattern $\lbrace \mathbf{x}(s_1,s_2)\rbrace$ with $s_1=0,\ldots,l_1-1$, $s_2=0,\ldots,l_2-1$ and components $x_i(s_1,s_2),~i=1,\ldots,d$, each consisting of $n$ observations. In the following, each component $x_i(s_1,s_2)$ is assumed to be corrected by its mean. The auto- and cross-periodograms for components $i$ and $j$ result from the DFT of the observed measurements,
 \begin{equation}\label{eq:DFTLattice}
\mathcal{F}_{i}(p,q)=\frac{1}{\sqrt{l_1l_2}}\sum_{s_1=0}^{l_1-1}\sum_{s_2=0}^{l_2-1}x_i(s_1,s_2)\exp\left[-2\pi\imath\left(\frac{ps_1}{l_1}+\frac{qs_2}{l_2}\right)\right]
\end{equation}
with $p=0,\ldots, l_1-1$ and $q=0,\ldots,l_2-1$ (cf. \cite{Renshaw1983}). 
From this expression, the auto-periodogram itself for frequencies $\mathbf{w}=(2\pi p/l_1,2\pi q/l_2)$ is obtained as 
\begin{eqnarray}\label{eq:autoperiodogramLattice}
\widehat{f}_{ii}(\mathbf{w})&=&\mathcal{F}_i(p,q)\overline{\mathcal{F}}_i(p,q)\\
&=& \lbrace a_i(p,q)\rbrace^2+\lbrace b_i(p,q)
\rbrace^2.\nonumber
\end{eqnarray}
Notice that, as pointed out by \citet{hannan1970} and \cite{Ripley1981}, we have $f_{ii}(0,0)=0$ as the periodogram is calculated using mean-corrected observations. Analogous to Section \ref{sec:DFTspp}, the cross-periodogram is obtained as $\widehat{f}_{ij}(\mathbf{w})=F_i(p,q)\overline{F}_j(p,q)$.

We remark that, apart from the calculation through demeaned observations, both sample spectra could also be computed through the DFT of the sample auto- and cross-covariance functions. However, while the sample auto- and cross-covariance functions themselves could be affected by auto-dependencies and the calculations of both sample spectra might be highly time consuming, the computation through demeaned observations is far quicker to evaluate  and less affected to round-off errors (see \cite{Renshaw1983}, \cite{Renshaw2002}).

Since $\widehat{f}_{ii}(w_{l_1-p}, w_q)=\widehat{f}_{ii}( w_p, w_{l_2-q})$, a reasonable form to output the periodogram is a matrix of dimensions $p=0,\ldots, l_1/ 2$ and $q=-l_2/ 2,\ldots,(l_2-1)/ 2$  such that it suffices to compute the periodogram over both negative and positive integers for one of the frequency coordinates and only over positive integers for the other  coordinate (cf. \cite{Renshaw1983}). 

We note that, as the observations are evaluated over integer values only, the highest row $(p)$ and column $(q)$ values that can be resolved (the Nyquist frequencies) are $p=(l_1-1)/2$ and $q=(l_2-1)/2$. This implies that any variability of higher, unresolvable frequencies is forced into lower frequencies such that the periodogram is affected by aliasing. That is, for integer values $s_1$ and $s_2$, no distinction between $\exp(-\imath( w_ps_1+ w_qs_2))$ and $\exp(-\imath(( w_p+2k\pi)s_1+( w_q+2k\pi)s_2))$ can be made (cf. \cite{Renshaw1983}, \cite{Mugglestone1990} and \cite{Kanaan2000}).

\section{Multivariate spatial hybrid processes}

While point and lattice processes have been treated separately in previous sections, this section covers the joint analysis of both types of spatial processes, where the point locations are assumed to coincide with the spatial lattice $\mathbf{S}_L$.

\subsection{Mixed spatial lattice-point processes: a spatial hybrid process}

Taking \cite{Kanaan2000} and \cite{Kanaan2008} as fundamental references, a spatial hybrid process is characterised as follows. Let $\Xi(\mathbf{s})=\left(N, X(\mathbf{s})\right)\T$  denote a bivariate spatial hybrid process with point component  $ N$ and lattice component $ X(\mathbf{s})$, respectively. In the remainder of this section,   $\Xi(\mathbf{s})$ is assumed to be stationary which implies that both component processes are jointly and marginally stationary. For the components, we additionally assume $N$ to be orderly and that $X(\mathbf{s})$ is corrected for its mean. 

\subsubsection{Second-order properties of spatial hybrid  processes}

Analogous to spatial point and lattice processes, a spatial hybrid process can be characterised by its first- and second-order moments. As the point- and lattice-type characteristics have been studied individually in the previous sections, none of these characteristics  will be redescribed in detail here. Besides these point- and lattice-type first- and second-order characteristics, cross-type characteristics are needed to explore structural interrelations between the point and the lattice components.  Assuming that the limit as $\nu(d\mathbf{s})\rightarrow 0$ exists, the cross-covariance density function is defined as
\begin{equation}
\label{eq:corsscovHybrid}
\zeta_{NX}(\mathbf{s,s}^\prime)=\lim\limits_{\nu(d\mathbf{s})\to 0} \left(\frac{\e\left[(N(d\mathbf{s})-\e\left[N(d\mathbf{s})\right])(X(\mathbf{s}^\prime)-\e\left[X(\mathbf{s}^\prime)\right])\right]}{\nu(d\mathbf{s})}\right)
\end{equation}

where $\zeta_{NX}(\mathbf{s,s}^\prime)=\zeta_{XN}(\mathbf{s}^\prime,\mathbf{s})$.

Under stationarity of $\Xi(\mathbf{s})$ \eqref{eq:corsscovHybrid} simplifies as follows. Writing $\mathbf{a}=\mathbf{s}-\mathbf{s}^\prime$, $\mathbf{b}=\mathbf{a}+\mathbf{c}$ and assuming that the lattice component is corrected for its mean we have   
\begin{equation}
\begin{aligned}
\zeta_{NX}(\mathbf{a})&= \lim\limits_{\nu(d\mathbf{a})\to 0} \left(\frac{\e\left[(N(d\mathbf{b})-\e\left[N(d\mathbf{b})\right])(X(\mathbf{c})-\e\left[X(\mathbf{c})\right])\right]}{\nu(d\mathbf{a})}\right)\\
&= \lim\limits_{\nu(d\mathbf{a})\to 0} \left(\frac{\e\left[N(d\mathbf{b})X(\mathbf{c})\right]}{\nu(d\mathbf{a})}\right)-\lambda\e\left[X\right]\\
&=\lim\limits_{\nu(d\mathbf{a})\to 0} \left(\frac{\e\left[N(d\mathbf{b})X(\mathbf{c})\right]}{\nu(d\mathbf{a})}\right).
\end{aligned}
\end{equation}
Notice that under stationarity of $\Xi(\mathbf{s})$, we also have $\zeta_{NX}(\mathbf{a})=\zeta_{XN}(-\mathbf{a})$.

Recapitulating the above results, a bivariate spatial hybrid process is said to exhibit complete spatial randomness if the point component is a homogeneous Poisson process and the lattice component is a Gaussian white noise which implies that $\zeta_{NX}(\cdot)=0$.

\subsubsection{Cross-spectral properties of spatial hybrid processes}

This section discusses the properties of the cross-spectral density function $f_{NX}(\mathbf{w})$ for a stationary spatial hybrid process $\Xi(\mathbf{s})$ which is, 
analogous with the previous sections, defined as the Fourier transform of the cross-covariance density function $\zeta_{NX}(\mathbf{a})$, 
\begin{equation}\label{EQ:hybrid:DTF}
f_{NX}(\mathbf{w})=\int \zeta_{NX}(\mathbf{a})\exp(-\imath\mathbf{w}\mathbf{a}\T)d\mathbf{a}.
\end{equation} 
Since $\zeta_{NX}(\mathbf{a})=\zeta_{XN}(\mathbf{-a})$ under stationarity of $\Xi(\mathbf{s})$, we have $f_{NX}(\mathbf{w})=f_{XN}(-\mathbf{w})$ and it suffices to compute only one cross-spectral density function. As for the point process case, the cross-spectrum is a complex-valued function 
and can be decomposed into either the co-spectrum $C_{NX}(\mathbf{w})$ and the quadrature spectrum $Q_{NX}(\mathbf{w})$ using Cartesian coordinates or the cross-amplitude spectrum $\mathfrak{a}_{NX}(\w)$ and the cross-phase spectrum $\wp_{NX}(\w)$ using polar coordinates. 

In the motion-invariant case, \eqref{EQ:hybrid:DTF} simplifies to 
\begin{equation}
\label{eq:isotropichybridspectra}
f_{NX}(\varpi)=2\pi \int_0^\infty r\zeta_{NX}(r)\J_0(r\varpi)dr
\end{equation}
where $\J_0(r\varpi)$ is the unmodified Bessel function of the first kind of order zero, defined in Section \ref{sec:sppSpectraTheo}, and $\varpi=\sqrt{ w_1^2+ w_2^2}$. In this particular case, as $f_{NX}(\varpi)$ is a real number, we have  $f_{NX}(\varpi)=C_{NX}(\varpi)=\mathfrak{a}_{NX}(\varpi)$ while  the quadrature spectrum $Q_{NX}(\varpi)$ and the cross-phase spectrum $\wp_{NX}(\varpi)$ are identically zero at all frequencies.

Notice that if the spatial hybrid process exhibits complete spatial randomness, all cross-spectral characteristics as well as the coherence spectrum are identically zero at all frequencies except the cross-phase spectrum which is undefined.

\subsubsection{Estimation of the cross-spectral properties of spatial hybrid processes}

Before we discuss spectral density functions for multivariate spatial hybrid processes and their representation as a spatial dependence graph model, we now concern the estimation of the cross-periodogram from a spatial bivariate hybrid pattern.

First, bivariate mixtures of regularly-shaped lattice and unmarked point processes are considered. Assume we have observed a lattice and a point pattern within a congruent rectangular region $\mathbf{S}\subset\mathds{R}^2$ with sides of lengths $l_1$ and $l_2$. Let $\lbrace\mathbf{s}_i\rbrace=\lbrace(x_i,y_i)\rbrace, i=1,\ldots,n$ denote the point locations and $\lbrace x(s_1, s_2)\rbrace$ denote the observed measurements recorded along a regular grid of size $s_1=0,\ldots,l_1-1$ and $s_2=0,\ldots,l_2-1$. Throughout this section, the  lattice component is assumed to be corrected for its mean and the point locations are assumed to be scaled to the unit square prior to the analysis.  

Using the previous results, the cross-periodogram follows as $\widehat{f}_{NX}(\mathbf{w})=\mathcal{F}_N(p,q)\overline{\mathcal{F}}_X(p,q)$ where  $\mathcal{F}_N(p,q)$ and $\mathcal{F}_X(p,q)$ are defined as in \eqref{eq:sppShortform} and \eqref{eq:DFTLattice}, respectively. Thus, we have 
\begin{equation}\label{eq:sampleCrosSpecHybrid}
\begin{aligned}
\widehat{f}_{NX}(\mathbf{w})= &\left(\sum^{n}_{i=1}\exp(-2\pi\imath(px_i+qy_i))\right) \times \\
&\left(\frac{1}{\sqrt{l_1l_2}}\sum_{s_1=0}^{l_1-1}\sum_{s_2=0}^{l_2-1}x(s_1, s_2)\exp\left[2\pi\imath\left(\frac{\bar{p}s_1}{l_1}+\frac{\bar{q}s_2}{l_2}\right)\right]\right)
\end{aligned}
\end{equation}
where $p = 0, 1, \ldots, 16$, $q = - 16, \ldots, 15$,  $\bar{p}=0,\ldots, l_1/ 2$ and $\bar{q}=-l_2/ 2,\ldots,(l_2-1)/ 2$.

Next, bivariate mixtures of irregularly-shaped spatial lattice and point patterns are considered. For the point component, consider we have observed a set of $n_P$ point locations $\mathbf{s}_i=(x_i,y_i), i=1,\ldots, n_P$. Similarly, let  $\mathbf{s}_j=(x_j,y_j), j=1,\ldots, n_L$ denote the set of coordinates computed from the centroids
of $n_L$ irregularly-shaped lattice entities. Notice that this approach also allows for regularly-shaped spatial lattice processes by taking the centroids of $n_L$ regularly-shaped grid squares into account. Then, the cross-periodogram for components $i$ and $j$  is obtained by substituting \eqref{ed:DFTmark} for $F_X(p,q)$ in  \eqref{eq:sampleCrosSpecHybrid} yielding 
\begin{equation}\label{eq:HYdrinIrregLatt}
\begin{aligned}
\widehat{f}_{ij}(\mathbf{w})= &\left(\sum^{n_P}_{i=1}\exp(-2\pi\imath(px_i+qy_i))\right) \times \\
&\left(\sum_{j=1}^{n_L}\left(m_j(\mathbf{s}_j)-\mu_M(\mathbf{s}_j)\right)\exp(2\pi\imath(px_j+qy_j))\right).
\end{aligned}
\end{equation}

As previously stated, \eqref{eq:HYdrinIrregLatt} can also been understood as a special case of a cross-spectral density function for a MMSPP where $\left(m_i(\mathbf{s}_i)-\mu_M(\mathbf{s}_i)\right)$ is set to $1$ for the unmarked point pattern. 

\subsection{Multivariate spatial hybrid processes}\label{MVAward}

In order to discuss partial interrelations within the context of spatial hybrid processes and to extend the spatial dependence graph model, we now cover possible extensions to multivariate hybrid processes. To begin, let $\boldsymbol{\Xi}(\mathbf{s})=(\mathbf{N}, \mathbf{X}(\mathbf{s}))\T$ denote a $d$-variate spatial hybrid process consisting of a multivariate spatial point process $\mathbf{N}$ with components $ N_i, i=1,\ldots, d_N$ and a multivariate spatial lattice process $\mathbf{X}(\mathbf{s})$ with components $ X_i(\mathbf{s}), i=1,\ldots, d_X$ where $d=d_N + d_X$. Adopting the former results, the lattice components are assumed to be corrected by their means and the point components are required to be orderly. In general, the number of components in $\mathbf{N}$ and $\mathbf{X}(\mathbf{s})$ is allowed to differ. However, in the following we assume that at least two components of both $\mathbf{N}$ and $\mathbf{X}(\mathbf{s})$ are contained in $\boldsymbol{\Xi}(\mathbf{s})$. 

To start, mixtures of multivariate regularly-shaped spatial lattice and multivariate spatial point processes are concerned. Under the usual assumptions, the auto-spectral (resp. cross-spectral) density function can be defined as the Fourier transform of the auto-covariance (resp. cross-covariance) density function. However, in contrast to the previous notions, we now consider cross-covariance density and cross-spectral density functions between similar and different types of spatial processes yielding different expressions for the cross-covariance and the corresponding cross-spectral density functions depending on the selected components of $\boldsymbol{\Xi}(\mathbf{s})$. For example, for the cross-covariance density functions we have: (a) $\zeta_{N_iN_j}(\cdot)$ for point-point cross-covariance density functions, (b) $\zeta_{X_iX_j}(\cdot)$ for lattice-lattice cross-covariance density functions and, finally, (c) $\zeta_{N_iX_j}(\cdot)$ for point-lattice cross-covariance density functions. The corresponding cross-spectral density functions for components $i$ and $j$ of $\boldsymbol{\Xi}(\mathbf{s})$ at frequencies $\mathbf{w}$ follow, under the usual assumptions, as the Fourier transform of either $\zeta_{N_iN_j}(\mathbf{c})$, $\zeta_{X_iX_j}(\mathbf{c})$ or $\zeta_{N_iX_j}(\mathbf{c})$ and could be estimated by means of point-point, lattice-lattice or point-lattice cross-periodograms, namely 
\begin{equation}
\begin{aligned}
\widehat{f}_{N_iN_j}(\mathbf{w})&=\mathcal{F}_{N_i}(p,q)\overline{\mathcal{F}}_{N_j}(p,q)\nonumber\\
\widehat{f}_{N_iX_j}(\mathbf{w})&=\mathcal{F}_{N_i}(p,q)\overline{\mathcal{F}}_{X_j}(p,q)\nonumber\\
\widehat{f}_{X_iX_j}(\mathbf{w})&=\mathcal{F}_{X_i}(p,q)\overline{\mathcal{F}}_{X_j}(p,q).\nonumber
\end{aligned}
\end{equation}
Here, $\mathcal{F}_{N_i}(p,q)$ is the discrete Fourier transform of \eqref{eq:sppShortform} and $\mathcal{F}_{X_i}(p,q)$ is as \eqref{eq:DFTLattice}, where we assume that the points have been scaled to the unit square. Likewise, depending on whether interrelations of similar or dissimilar components are considered, three different spectral coherence functions can be considered.
 
Next, mixtures of multivariate irregularly-shaped spatial lattice data and multivariate spatial point processes are of interest where $\mathbf{s}$ either refers to the set of $n_{d_N}$ point locations or the set of $n_{d_X}$ coordinates representing the centroids
of irregularly-shaped lattice entities. As previously mentioned, this also covers multivariate regularly-shaped spatial lattice processes recorded at centroids of grid squares. For such multivariate spatial hybrid processes, we can model the multivariate irregularly-shaped spatial lattice processes by means of a MMSPP. As before, estimates for all cross-spectral density functions of $\boldsymbol{\Xi}(\mathbf{s})$ could be obtained by means of cross-periodograms at frequencies $\mathbf{w}$ where the point-lattice cross-periodogram $\widehat{f}_{N_iX_j}(\mathbf{w})$ and the lattice-lattice cross-periodogram $\widehat{f}_{X_iX_j}(\mathbf{w})$ are defined by 
\begin{equation}\label{eq:hybridNZnew}
\begin{aligned}
\widehat{f}_{N_iX_j}(\mathbf{w})= &\left(\sum^{n_P}_{i=1}\exp(-2\pi\imath(px_i+qy_i))\right) \times \\
&\left(\sum_{j=1}^{n_L}\left(m_j(\mathbf{s}_j)-\mu_M(\mathbf{s}_j)\right)\exp(2\pi\imath(px_j+qy_j))\right)\\
\end{aligned}
\end{equation}
and 
\begin{equation}\label{eq:hybridZZasMMSPP}
\begin{aligned}
\widehat{f}_{X_iX_j}(\mathbf{w})= &\left(\sum_{i=1}^{n_P}\left(m_i(\mathbf{s}_i)-\mu_M(\mathbf{s}_i)\right)\exp(-2\pi\imath(px_i+qy_i))\right) \times \\
&\left(\sum_{j=1}^{n_L}\left(m_j(\mathbf{s}_j)-\mu_M(\mathbf{s}_j)\right)\exp(2\pi\imath(px_j+qy_j))\right).
\end{aligned}
\end{equation}

\subsection{Spatial dependence graph model for multivariate hybrid data}

This section extends the spatial dependence graph formalism introduced in the papers of \cite{Eckardt2016} for multivariate and \cite{EckardtMateuSDGM} for multivariate-marked point processes to the present context. Adopting the results of these papers, a mixed-type spatial dependence graph model (mSGDM) is defined as an undirected graph $\G=(\V,\E)$ with vertex set $\V$ and edge set $\E$ in which missing edges depict conditional independence between the components of $\boldsymbol{\Xi}(\mathbf{s})$ which could either be of lattice or of point process nature. 

To this end, let $\Xi_i(\mathbf{s})$ and $\Xi_j(\mathbf{s})$ denote the $i$-th and the $j$-th components of $\boldsymbol{\Xi}(\mathbf{s})$, respectively, and $ \Xi_{\V\backslash\lbrace i,j\rbrace}$ be the set of all alternative components contained in $\boldsymbol{\Xi}(\mathbf{s})$. Associating each of the $d$ components with a vertex of the SDGM, the following relation holds  
 \begin{equation*}
\condindep{\Xi_i(\mathbf{s})}{\Xi_j(\mathbf{s})}{ \Xi_{\V\backslash\lbrace i,j\rbrace}((\mathbf{s})}\Leftarrow \lbrace v_i,v_j\rbrace\notin\E
\end{equation*}    
where $\E=\lbrace \lbrace v_i,v_j\rbrace: R_{ij\given \V\backslash\lbrace i,j\rbrace}(\mathbf{w})\neq 0\rbrace$ and $R_{ij\given \V\backslash\lbrace i,j\rbrace}(\mathbf{w})$ is the partial spectral coherence function. Notice that, as both point and lattice components are considered, this conditional independence relation includes the following statements:
 \begin{equation*}
 \begin{aligned}
 \condindep{ N_i}{ N_j}{\Xi_{\V\backslash\lbrace i,j\rbrace}(\mathbf{s})}&\Leftarrow \lbrace v_i,v_j\rbrace\notin\E\\
  \condindep{ X_i(\mathbf{s})}{ X_j(\mathbf{s})}{\Xi_{\V\backslash\lbrace i,j\rbrace}(\mathbf{s})}&\Leftarrow \lbrace v_i,v_j\rbrace\notin\E\\
   \condindep{ N_i}{ X_j(\mathbf{s})}{\Xi_{\V\backslash\lbrace i,j\rbrace}(\mathbf{s})}&\Leftarrow \lbrace v_i,v_j\rbrace\notin\E.
\end{aligned}
\end{equation*}   
As discussed by \cite{Eckardt2016} and \cite{EckardtMateuSDGM},   the mixed-type SDGM can be computed from the partial cross-spectral density, partial spectral coherence or absolute rescaled inverse spectral density functions.

\subsection{General formalism for multivariate spatial data}

We now propose a general framework which covers any type of spatial data in a unified approach. To this end, let $\boldsymbol{\Theta}(\mathbf{s}_n)$ denote a multivariate  spatial process consisting of $d$ generic components $ \Theta_i(\mathbf{s}_n),~i=1,\ldots,d$  which could either be of geostatistical, spatial lattice or spatial point process nature. For both, regularly- and irregularly-shaped spatial lattice processes, we assume that the observations have been recorded at centroids, either of polygon entities or grid squares, such that both spatial lattice and geostatistical processes coincide. Besides, any lattice or geostatistical component is assumed to be corrected by its means. For any point process contained in $\boldsymbol{\Theta}(\mathbf{s}_n)$, we assume orderliness and that the spatial point patterns have been scaled to the unit square prior to the analysis. 

We note that, analogous to \eqref{eq:hybridZZasMMSPP}, any spatial auto- or cross-spectral density function can be treated as auto- or cross-spectral density of a MMSPP, e.g. by considering the observed values as a quantitative mark of the centroids or by setting the difference $\left(m_i(\mathbf{s}_i)-\mu_M(\mathbf{s}_i)\right)$ to one in case of multivariate point patterns. Consequently, the definition of a general graphical model coincides with the definition of the mSDGM for MMSPP.

Using the results of the previous sections, a general SDGM is defined as follows. Let $ \Theta_i(\mathbf{s})$ and $ \Theta_j(\mathbf{s})$ denote the $i$-th and the $j$-th components of $\boldsymbol{\Theta}(\mathbf{s})$, respectively, and $ \Theta_{\V\backslash\lbrace i,j\rbrace}$ be the set of all alternative components contained in $\boldsymbol{\Theta}(\mathbf{s})$. Associating each of the $d$ components with a vertex of the SDGM, the following relation holds  
 \begin{equation*}
\condindep{\Theta_i(\mathbf{s})}{\Theta_j(\mathbf{s})}{\Theta_{\V\backslash\lbrace i,j\rbrace}(\mathbf{s})}\Leftarrow \lbrace v_i,v_j\rbrace\notin\E
\end{equation*}    
where $\E=\lbrace \lbrace v_i,v_j\rbrace: R_{ij\given \V\backslash\lbrace i,j\rbrace}(\mathbf{w})\neq 0\rbrace$ and $R_{ij\given \V\backslash\lbrace i,j\rbrace}(\mathbf{w})$ is the partial spectral coherence function computed from $\boldsymbol{\Theta}(\mathbf{s}_n)$. 

Adopting the result of \cite{Eckardt2018partial}, we note that these partial spectra characteristics can then, in turn, also be used to define partial spatial characteristics from any type of multivariate spatial data. 

\section{Application}\label{sec:2}

This section illustrates the application of the proposed graphical model using multivariate hybrid data on point locations for eleven pre-classified crime categories at street-level and aggregated ambulance service call-out incidents at ward-level recorded in London. Both datasets were collected over a one-month period in December 2015 and have been made available under the Open Government Licence by the British Home Office for London. 

The areal data on aggregated ambulance service call-out incidents was downloaded from \url{https://data.london.gov.uk/dataset/} and provides information on the numbers of incidents of assaults (including assaults against women and teens), binge drinking (meaning alcohol poisoning), injuries caused by any type of weapon, cocaine overdose, and heroin overdose at ward-level. Records were available for $599$ of $607$ wards for London and reported either aggregated numbers for incidents or zeros if no incidents occurred. Relevant information was collected and classified by the London Ambulance Service by inspecting different sources based on records of all ambulances despatched in London. Incident cases for assault, the usage of weapons, and the appearance of alcohol related illnesses were derived from retrospective records by paramedics and ambulance staff. Records on alcohol related illnesses were relabelled as binge drinking for the subset of patients aged forty or younger. Finally, information on the type of drugs and the type of weapons originated from notes by the emergency telephone number handler. 

For the point components, we consider open data on crimes which has been downloaded from \url{https://data.police.uk/data/}. This data contains pairs of coordinates for different crime categories at street-level, either within a 1 mile radius of a single point or within a custom area of a street. The crime categories were generated by local officials. For our analysis we pre-selected a subset of $11$  out of $14$ crime categories.

\subsection{Point and lattice characteristics computed from the point and lattice components}

To provide a first impression of both datasets, different descriptive statistics are discussed first. For the ambulance service call-out data, we calculated the median ($\tau_L$), the mean ($\mu_L$), the first and third quantiles from the original data where we excluded any zero cases prior to the computation (see Table \ref{tab:hybridAmbulanceSummaryCases}).
\begin{table}
\caption{\label{tab:hybridAmbulanceSummaryCases}Summary statistics computed from the ambulance service call-out data, zero cases excluded}
\fbox{%
\begin{tabular}{*{9}{c}}
\em incident type&\em $n$&\em $n_W$&\em $min$&\em $Q_{25}$&\em$\tau_L$&\em $\mu_L$&\em $Q_{75}$&\em $max$\\
\hline
  Assault & 1465 & 476 & 1 & 1 & 2 & 3.078 & 4 & 24 \\ 
  Binge Drinking & 3070 & 559 & 1 & 2 & 4 & 5.492 & 6 & 120 \\ 
  Cocaine overdose & 64 & 28 & 2 & 2 & 2 & 2.286 & 2 & 4 \\ 
  Heroin overdose & 82 & 34 & 2 & 2 & 2 & 2.412 & 2 & 6 \\ 
  Injuries (all weapons) & 245 & 197 & 1 & 1 & 1 & 1.244 & 1 & 4 \\ 
 \end{tabular}}
\end{table}

Inspecting this table, we found that binge drinking was reported most frequently whereas the lowest numbers appeared for cocaine and heroin overdose.  Further, at least one case of binge drinking was recorded for all $599$ wards ($n_W$). Different from this, cocaine and heroin overdose were only reported for $28$ and $34$ wards, respectively.   

Next, different point process characteristics computed from the London crime data are discussed. Inspecting the numerical summary statistics computed from this data (see Table \ref{tab:hybridCrimeSummary}), we observed that anti-social behaviour appeared most frequently. Further, as all $CEI$ values are all below the threshold value of  $1$, all patterns are to be considered as clustered. 
\begin{table}
\caption{\label{tab:hybridCrimeSummary}Numerical summary characteristics computed from the London crime data}
\fbox{%
\begin{tabular}{*{8}{c}}
\em crime type&\em $n$&\em prop. $(\%)$&\em $\lambda$&\em $\mu_D$&\em $\tau_D$ &\em $IQR_D$ &\em CEI\\
\hline
Anti-social behaviour & 8963 & 31.967 & 28854.541 & 0.002 & 0.002 & 0.001 & 0.685  \\ 
Bicycle theft &  643 & 2.293 & 2070.007 & 0.006 & 0.004 & 0.005 & 0.559 \\ 
Burglary & 4277 & 15.254 & 13768.925 & 0.003 & 0.002 & 0.002 & 0.663  \\ 
Criminal damage and arson & 2843 & 10.140 & 9152.456 & 0.003 & 0.003 & 0.003 & 0.666 \\ 
Possession of weapons &  124 & 0.442 & 399.193 & 0.016 & 0.013 & 0.017 & 0.637 \\ 
Public order & 1345 & 4.797 & 4329.952 & 0.005 & 0.004 & 0.004 & 0.657  \\ 
Robbery &  725 & 2.586 & 2333.989 & 0.007 & 0.005 & 0.005 & 0.630 \\ 
Shoplifting &  592 & 2.111 & 1905.823 & 0.006 & 0.004 & 0.007 & 0.551 \\ 
Theft from the person &  719 & 2.564 & 2314.673 & 0.006 & 0.004 & 0.005 & 0.552 \\ 
Vehicle crime & 3262 & 11.634 & 10501.340 & 0.003 & 0.003 & 0.003 & 0.674 \\ 
Violence and sexual offences & 4545 & 16.210 & 14631.696 & 0.003 & 0.002 & 0.002 & 0.674 \\
\end{tabular}}
\end{table}

To compare the London crime and ambulance service call-out data and to allow for a joint analysis of the spatial hybrid data by means of classical multivariate techniques, we aggregated the point locations of the London crime data at ward level and considered the crime counts per ward as inputs for different lattice type characteristics and calculated the median ($\tau_L$), the mean ($\mu_L$) and the first and third quantiles based on non-zero cases only (see Table \ref{tab:hybridLondonLAtticeNonZeros}). Looking at this table, a great variability among the different types of crimes at ward-level can be observed.  

\begin{table}
\caption{\label{tab:hybridLondonLAtticeNonZeros}Summary statistics computed from the aggregated crime data, zero cases excluded}
\fbox{%
\begin{tabular}{*{9}{c}}
\em incident type&\em $n$&\em $n_W$&\em $min$&\em $Q_{25}$&\em$\tau_L$&\em $\mu_L$&\em $Q_{75}$&\em $max$\\
\hline
Anti-social behaviour & 8963 & 606 & 1 & 9 & 13 & 14.79 & 18 & 100 \\ 
Bicycle theft & 643 & 311 & 1 & 1 & 1 & 2.07 & 3 & 11 \\ 
Burglary & 4277 & 598 & 1 & 4 & 6 & 7.15 & 9 & 34 \\ 
Criminal damage and arson & 2843 & 588 & 1 & 3 & 4 & 4.84 & 6 & 20 \\ 
Possession of weapons & 124 & 106 & 1 & 1 & 1 & 1.17 & 1 & 3 \\ 
Public order & 1345 & 498 & 1 & 1 & 2 & 2.7 & 4 & 17 \\ 
Robbery & 725 & 361 & 1 & 1 & 2 & 2.01 & 2 & 16 \\ 
Shoplifting & 592 & 315 & 1 & 1 & 1 & 1.88 & 2 & 17 \\ 
Theft from the person & 719 & 319 & 1 & 1 & 1 & 2.25 & 2 & 17 \\ 
Vehicle crime & 3262 & 592 & 1 & 3 & 5 & 5.51 & 7 & 20 \\ 
Violence and sexual offences & 4545 & 603 & 1 & 4.5 & 7 & 7.53 & 10 & 25 \\ 
\end{tabular}}
\end{table}

\subsection{Multivariate analysis of the hybrid data}\label{sec:MVAward}

This section discusses the results of the multivariate analysis computed from both types of spatial data contained in the London crime and ambulance service call-out data. To this end, we adopted the ideas of Chapter 4.9 of \cite{Illian2008} and considered different numerical summary characteristics as inputs for a hierarchical cluster analysis, a principal component analysis and parallel coordinates charts.
Starting with the results calculated from the lattice and the point components of the hybrid data, the findings of the joint analysis of both the lattice and the aggregated point components are presented.  

 For the lattice and aggregated point components, we considered the empirical mean ($\mu_L$),  range ($rg$), Moran's $I$ \citep{moran} and Geary's $C$ \citep{geary} as inputs for the multivariate analysis whereas estimates of the mean nearest neighbour distance ($\mu_D$), the median nearest neighbour distance ($\tau_D$), the interquartile range of nearest neighbour distances ($IQR_D$) and the Clark-Evans index \citep{clarkevans1954} ($CEI$) are considered as inputs for the point components. Both $\mu_D$ and $\mu_L$ as well as $\tau_D$, $IRQ_D$ and $rg$ were chosen to control for the distributional characteristics and the heterogeneity among the observations,  while both autocorrelation statistics and the $CEI$ were selected as univariate measures of spatial association among the observations.  

\begin{figure}[h!]
\centering
\makebox{\includegraphics[scale=0.76]{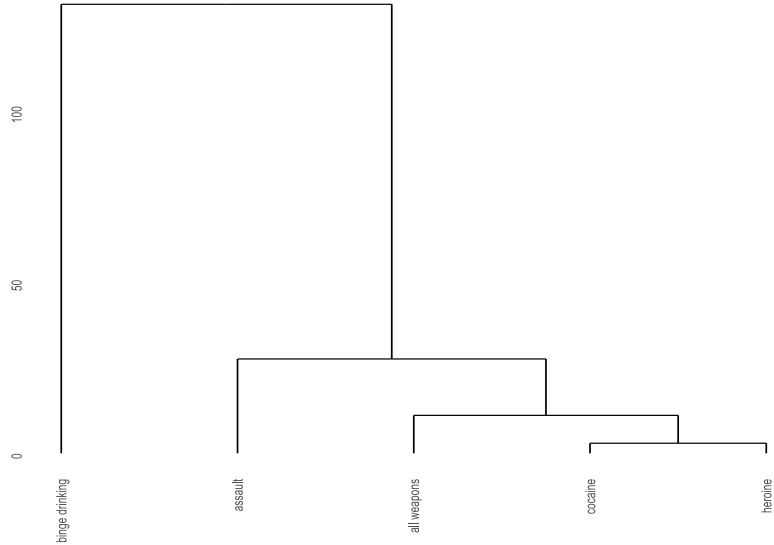}}
      \caption{Dendrogram computed from the lattice components of the spatial hybrid data from a hierarchical agglomerative cluster analysis using Ward's algorithm.\label{fig:dendogramWard1}}
\end{figure}

First, the results of the agglomerative hierarchical cluster analysis computed from both types of spatial data are presented. For the ambulance service call-out data at least two main clusters can be identified using Ward's algorithm. Reading off  the dendrogram for the ambulance service call-out data depicted in Figure \ref{fig:dendogramWard1}, cluster $1$ consists of four incidents (assault, cocaine overdose, heroin overdose, injuries (all weapons)) while cluster $2$ only consists of one incident (binge drinking). Reinspecting Table  \ref{tab:hybridAmbulanceSummaryCases}, a clear distinction between cluster $1$ and cluster $2$ can be made with respect to the summary statistics. While cluster $2$ is characterised by larger values for $\mu_L$, $max$, $\tau_L$, the highest numbers of incidents ($n$) appeared in almost all wards under study, no clear distinction between cluster $1$ and cluster $2$  can be made concerning both autocorrelation statistics. Considering the characteristics reported for cluster $1$, large differences of the summary characteristics between assault and the three alternative types of incidents can be observed. 

Turning to the results of the hierarchical cluster analysis computed from the London crime data, at least three main clusters can be identified using Ward's algorithm (see Figure \ref{fig:dendogramWard1}). Cluster $1$ consists of six types of crimes (anti-social behaviour, burglary, criminal damage and arson,   public order, vehicle crime, violence and sexual offences), cluster $2$ of two types of crimes (possession of weapons,  robbery) and cluster $3$ of three types of crimes (bicycle theft, shoplifting, theft from the person). Reconsidering the numerical summary characteristics reported in Table \ref{tab:hybridCrimeSummary} yields the following. A clear distinction can be made between cluster $1$ and the two alternative clusters. While cluster $1$ is characterised by the highest numbers of crime events and the smallest values for $\mu_D$, $\tau_D$ and $IRQ_D$, all alternative crimes appeared less frequently and occurred less close in terms of distances.  
\begin{figure}[h!]
\centering 
\makebox{\includegraphics[scale=0.8]{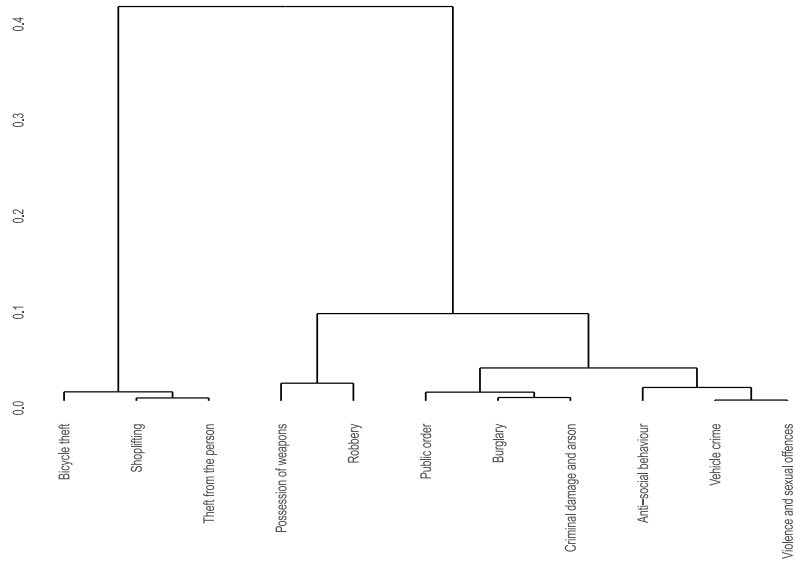}}
      \caption{Dendrogram computed from the point components of the spatial hybrid data from a hierarchical agglomerative cluster analysis using Ward's algorithm.\label{fig:dendogramWard2}}
\end{figure}

\begin{figure}
\centering
 \makebox{\includegraphics[scale=.6]{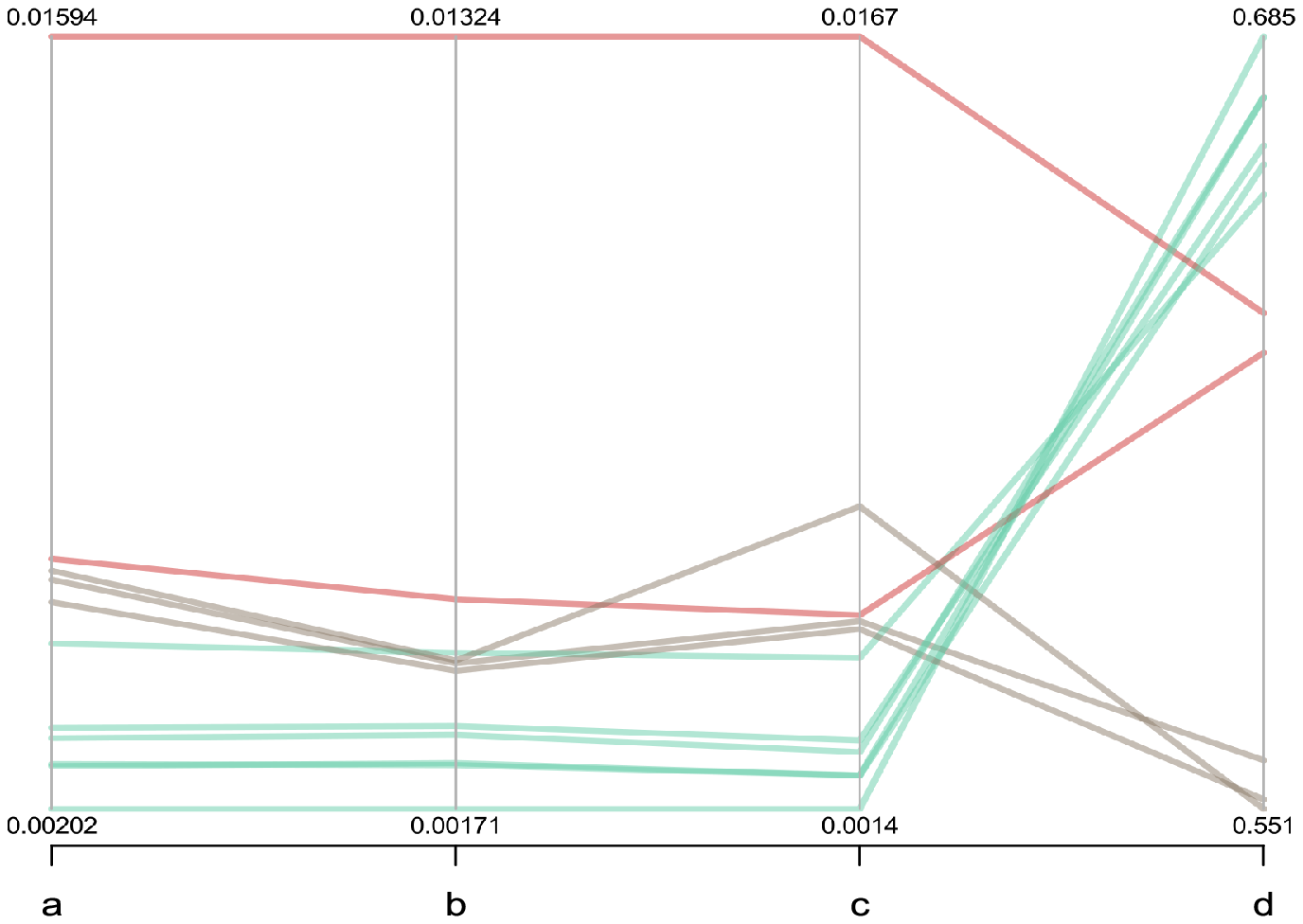}}
\caption{Parallel coordinates chart computed from the point components of the spatial hybrid data. Parallel axes are defined as follows: a = $\mu_D$, b = $\tau_D$, c = $IQR_D$ and d = $CEI$. Colours of lines depict the cluster membership obtained from the $3$-cluster solution using Ward's algorithm: cluster 1 (green), cluster 2 (red), and cluster 3 (brown).\label{Fig:parcoordhybridP2}}
\end{figure}

To investigate the characteristics of the $2$-cluster and the $3$-cluster solution for the ambulance service call-out data and the London crime data, parallel coordinates charts have been generated. 
\begin{figure}[h!]
\centering
 \makebox{\includegraphics[scale=.6]{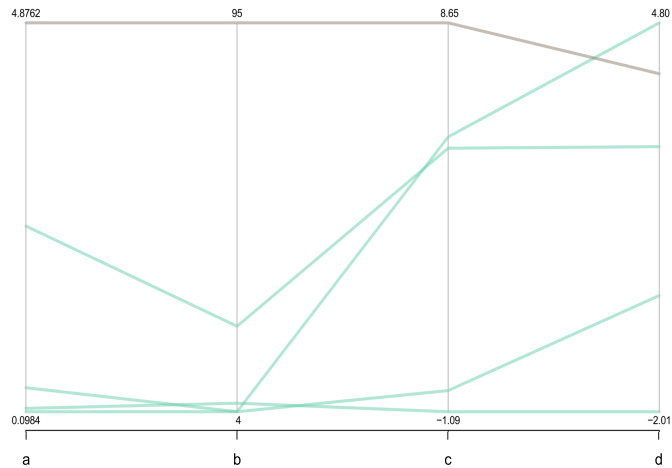}}
\caption{Parallel coordinates chart computed from the lattice components of the spatial hybrid data. Parallel axes are defined as follows: a = $\mu_L$, b = $rg$, c = Moran's $I$ and d = Geary's $C$. Colours of lines depict the cluster membership obtained from the $2$-cluster solution using Ward's algorithm: cluster 1 (green) and cluster 2 (brown). \label{Fig:parcoordwardP1}}
\end{figure} 
Inspecting the parallel coordinates chart for the lattice components depicted in Figure \ref{Fig:parcoordwardP1}, a clear separation of cluster $1$ (green) and cluster $2$ (brown) can be detected on the first three axes. For the fourth axis, however, an overlap of both clusters can be observed. Looking at the parallel coordinates chart for the point components shown in Figure \ref{Fig:parcoordhybridP2}, we observed that all three clusters are well separated on the $d$-axis, whereas no clear distinction can be made across the first three axes $a$ to $c$.    

We now discuss the results of the PCA computed from the lattice and the point components. For the lattice components, we found that the first two principal components explain $98.86\%$ of the variation. 

\begin{figure}[h!]
\centering
\makebox{\includegraphics[scale=0.6]{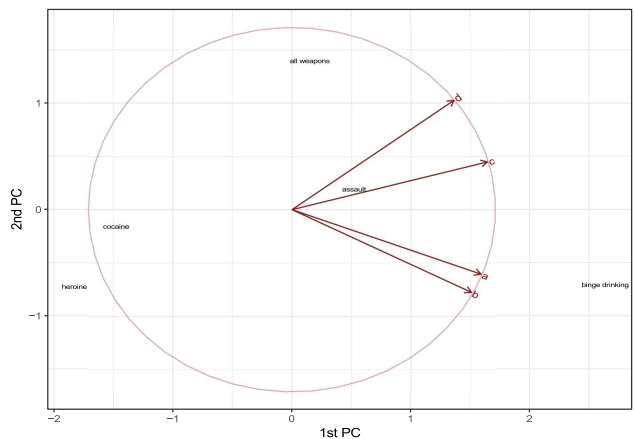}}
      \caption{\label{fig:biplotWard} Biplot of the PCA of the summary characteristics for the $5$ ambulance incidents on the first two principal components where $a=\mu_L$, $b=rg$, $c=I$ and $d=C$.}
\end{figure}

 Inspecting the loadings on the first two principal components, we found that all characteristics are positively associated with the first principal component while the two summary characteristics are negatively and the two autocorrelation statistics are positively associated with the second principal component. For the first principal component, the strongest loading is reported for Moran's $I$-statistic followed by $\mu_L$, $rg$ and Geary's $C$. However, as only small differences between the four loadings appeared, none of these characteristics dominated the first principal component. For the second principal component, we observed the strongest positive loading for Geary's $C$ and the strongest negative loading for $rg$. Reinspecting the values of all four loadings, we conclude that neither the summary characteristics nor the autocorrelation statistics dominated the second principal component. Inspecting the biplot of the PCA shown in Figure \ref{fig:biplotWard}, a clear separation of binge drinking from all remaining incidents can be observed. Besides, we found a close association of both drug related incidents.

 Applying the PCA to the London crime data, we observed that $99.66\%$ of the variation is explained by the first two principal components. Inspecting  the loading on the first two principal components, we found that the empirical mean, median  and interquartile range are positively  and the Clark-Evans index is negatively  associated with the first principal components, while all four numerical summary characteristics are negatively associated with the second principal component.
 For the first principal component,  the strongest positive loading appeared for $\mu_D$ followed by $IOR_D$ and $\tau_D$. As only small differences between the loadings appeared, we conclude that none of the four numerical summary characteristics dominated the first principal component. Reinspecting the loadings of the second principal component, we observed a high negative loading for the Clark-Evans index while only small negative loadings are reported for the three alternative numerical summary characteristics. This indicates  that the second principal component seems to be dominated by the $CEI$. Inspecting the biplot of the PCA, a clear separation of possession of weapons and all remaining crimes can be identified. Besides, we observed two groupings of closely associated crimes: (a) bicycle theft, shoplifting, theft from the person and (b) anti-social behaviour, burglary, criminal damage and arson, public order, vehicle crime, and violence and sexual offences.
\begin{figure}[h!]
\centering
\makebox{\includegraphics[scale=0.6]{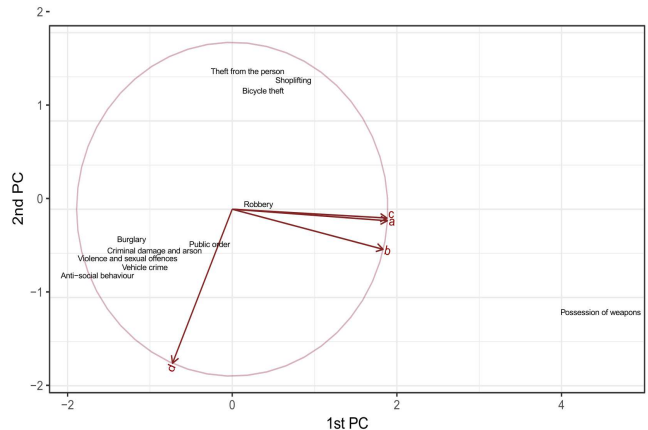}}
      \caption{\label{fig:biplotWardhybrid} Biplot of the PCA of the summary characteristics for the $11$ types of crime on the first two principal components where $a=\mu_D$, $b=\tau_D$, $c=IQR_D$ and $d=CEI$.}
\end{figure}

\begin{figure}[h!]
\centering
    \makebox{\includegraphics[scale=0.75]{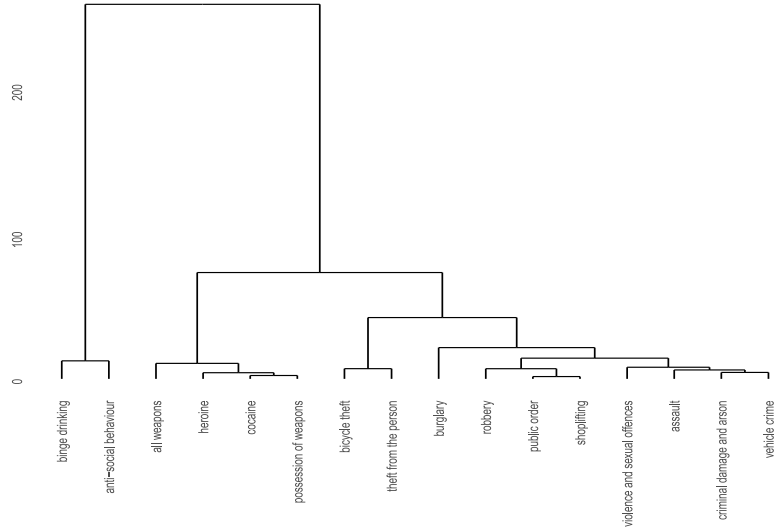}}
           \caption{Dendrogram computed from the aggregated point process and the lattice components of the London crime and ambulance service call-out data  from a hierarchical agglomerative cluster analysis using Ward's algorithm. \label{fig:dendogramWardLatticePoint}}
\end{figure}

We now turn to the result obtained from the joint analysis of the ambulance service call-out data and the aggregated crime counts. First, the results of the agglomerative hierarchical cluster analysis are presented. Inspecting the dendrogram of the cluster analysis (see Figure \ref{fig:dendogramWardLatticePoint}), at least three clusters can be identified using Ward's algorithm. Under this partitioning of the data, the first cluster consists of ten types of incidents (assault, bicycle theft, burglary, criminal damage and arson,  public order, robbery, shoplifting,  theft from the person, vehicle crime, violence and sexual offences), the second cluster of four types of incidents (cocaine overdose, heroin overdose, injuries (all weapons), possession of weapons) and the third cluster of two types of incidents (anti-social behaviour, binge drinking). To investigate the characteristics of these three clusters, parallel coordinates were calculated.

\begin{figure}[h!]
\centering
 \makebox{\includegraphics[scale=.6]{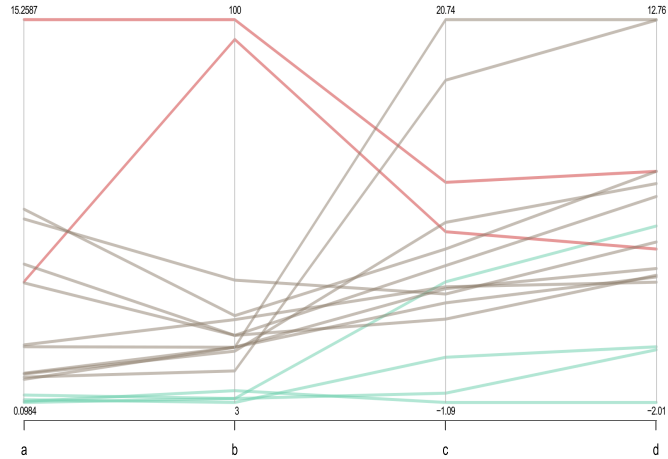}}
\caption{Parallel coordinates chart computed from the ambulance service call-out data and the aggregated crime count.  Parallel axes are defined as follows: a = $\mu_L$, b = $rg$, c = Moran's $I$ and d = Geary's $C$. Colours of lines depict the cluster membership obtained from the $3$-cluster solution using Ward's algorithm: cluster $1$ (brown), cluster $2$ (green) and cluster $3$ (red). \label{fig:parcoordhybridBOTHLatticePoints}}
\end{figure}

 Inspecting the parallel coordinates chart computed from the lattice-type characteristics (see Figure \ref{fig:parcoordhybridBOTHLatticePoints}), a clear distinction can be made on the $b$-axis between cluster $3$ and both alternative clusters.  
   
Finally, we present the result of the PCA on the ambulance service call-out data and the aggregated crime counts. Here, we observed that $93.35\%$ of the variation is explained by the first two principal components. Inspecting the loadings of the summary characteristics and the autocorrelation statistics on the first two principal components, we found that all four characteristics are positively associated with the first principal component. For the second principal component, we observed a positive association of the autocorrelation statistics and a negative association of the summary characteristics. 

\subsection{Joint analysis using the spatial dependence graph model}

We now present the results of the mSDGM computed from the marked partial spectral characteristics of the spatial hybrid data. Both the lattice and the point components were preprocessed as follows. For the lattice components, we computed the centroids for $599$ wards and attached the corresponding longitudes and latitudes to the data. Next, for each type of incident, we computed  demeaned values using global means calculated over all $599$ spatial sites. To each of these $599$ demeaned values, we attached the type of incident as qualitative mark.
Both, the  pair of coordinates and the qualitatively marked demeaned values were then rearranged in form of a multivariate-marked spatial point pattern where the longitudes and latitudes were considered as point locations and the demeaned incidents as quantitative mark.  For the point components, we attached a vector of ones to the data which served as an auxiliary quantitative mark. Finally, both datasets were matched into one dataset. 

\begin{figure}[h!]
\centering
\makebox{\includegraphics[scale=0.55]{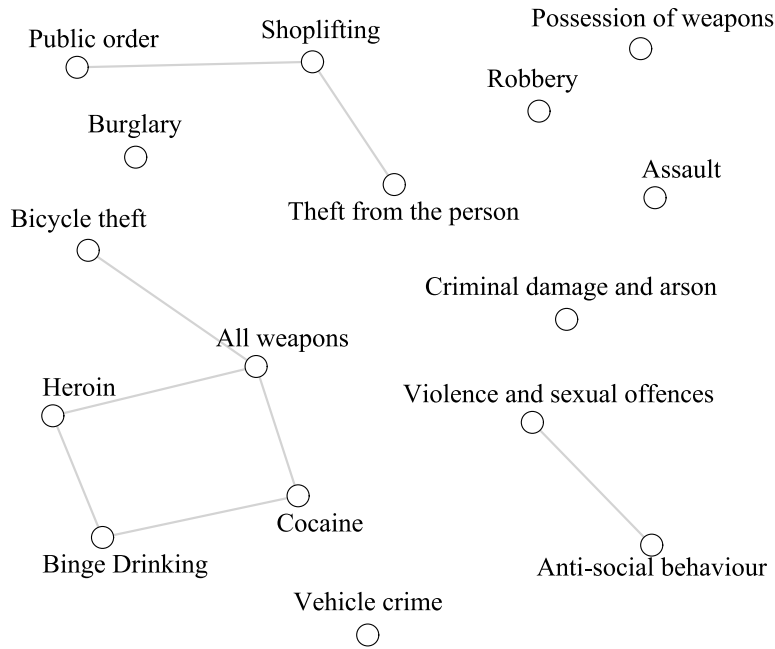}}
      \caption{\label{fig:hybrid03} Marked spatial dependence graph model for the London crime and ambulance service call-out data at ward-level, and the number of incidents as quantitative mark  for a threshold level of $\xi=0.3$.}
\end{figure}

To control for possible variations in strength of the partial interrelations between different pairs of incidents in the multivariate-marked data, we considered a threshold level of $\xi= 0.3$ in order to detect conditional partial interrelations with a weak effect size such that an edge is drawn between the nodes $i$ and $j$ if the supremum of the empirical  absolute rescaled inverse spectral density function for components $i$ and $j$ equals or exceeds $\xi$ for at least one frequency $\mathbf{w}$ for $p = 0, 1, \ldots, 16$ and $q = - 16, \ldots, 15$. That is, edges indicate that the strength of the linear partial interrelation between two component processes is greater than or equal to $\xi=0.3$.  In this particular case, the point distributions of the components $i$ and $j$ are said to be interrelated. The resulting mSDGM is  shown in Figure  \ref{fig:hybrid03}.

Inspecting this mSDGM, one pair of nodes (violence and sexual offences, anti-social behaviour), a $3$-node subgraph (public order, shoplifting, theft from the person), a $5$-node subgraph (bicycle theft, all weapons, heroin, cocaine, binge drinking) and $6$ isolated nodes (vehicle crime, assault, robbery, possession of weapons, burglary, criminal damage and arson) can be observed. We note that, except for the $5$-node subgraph, only interrelations of either the point or the lattice components can be detected in the mSDGM. 

For the isolated nodes, we conclude from the mSDGM that the marked spatial distributions of these types of incidents are not interrelated with the marked spatial distribution of any alternative type of incidents in the multivariate-marked data. This could indicate that the spatial distributions of the demeaned counts for these particular types of incidents are different from those of any alternative type of incidents in the London crime and ambulance service call-out data from a social, criminological or geographical perspective. However, reinspecting the numerical characteristics of the  isolated nodes, we found that both frequent and rare types of incidents are contained in this particular subset of nodes. Besides, we also observed that both incidents reported in the ambulance service call-out data and incidents reported in the London crime data are represented as isolated nodes. 

Reinspecting the $3$-node and the $5$-node subgraph structures of this mSDGM yields the following.   For the $3$-node subgraph, we found that the marked distributions of public order and theft from the person are indirectly interrelated through the marked distribution of shoplifting. This implies that both public order and theft from the person are independent given knowledge on the marked locations of shoplifting. Looking at the $5$-node subgraph, we observed that the marked locations of bicycle theft are conditionally independent from those of all alternative incidents given the marked locations of all weapons. Interestingly, no direct association can be detected between heroin overdose and cocaine overdose which are both jointly linked to binge drinking and injuries (all weapons).  This close relationship between heavy alcolhol  drinking, drug abuse and weapon injuries, especially self-inflicted firearm injuries, has also been reported by public health researchers and epidemiologists (cf. \cite{Branas2016},  \cite{Webster2009}, \cite{Wintemute2011}).

Comparing these results with the multivariate analysis presented in Section \ref{sec:MVAward} yields the following. Reinspecting the dendrogram computed from the lattice components shown in Figure \ref{fig:dendogramWard1}, we found that three incidents of cluster $1$ (injuries (all weapons), cocaine, heroin) are contained in the $5$-node subgraph. All these three incidents are most closely related to each other and also well separated from assault and binge drinking according to the dendrogram. While assault is not interrelated with any of these three incidents, binge drinking is interrelated to heroin and cocaine overdose and indirectly interrelated to injuries (all weapons) through either cocaine or heroin overdose. Looking at the dendrogram computed from the point components depicted in Figure  \ref{fig:dendogramWard2}, we found that two of the three crimes contained in cluster $3$ (namely shoplifting and theft from a person) are also connected in the mSDGM while bicycle theft, which is also contained in  cluster $3$, is not connected to any alternative point component. This close relationship between injuries (all weapons), cocaine and heroin overdose and also between public order and shoplifting can also be identified reading of the dendrogram computed from the joint analysis of both components. 

\section{Conclusions and discusion}

The growing availability and accessibility of multivariate spatial data and the rapid developments 
in geographical information systems (GIS) have led to an everincreasing
demand for statistical efficient methods that are able to account for the inherent complexity and structural 
interrelations of such data, while facilitating a clear interpretation. This paper contributes to the 
multivariate analysis of spatial data, presenting a unifying approach based
on partial marked point characteristics which allows for the simultaneous analysis of any type of 
multivariate spatial data by means of spatial undirected graphical models.

One main advantage of our approach is that it can handle, explore and analyse potential interrelations 
between different types of spatial data, for example, between different point- and lattice-type components 
in a multivariate setting. It is in this sense that we consider that our graphical approach presents a unifying strategy
for the overall analysis of multitype and multivariate spatial data.

We have analysed multivariate hybrid data on point locations for eleven pre-classified crime categories 
at street-level, and aggregated ambulance service call-out incidents at ward-level recorded in London. In particular, 
and using the mSDGM computed from the marked partial spectral characteristics of the spatial hybrid data, we can 
dissentangle partial interrelations between different pairs of incidents in the multivariate-marked data, and provide
information on conditional independence of any two types of crimes, given a third one.

We have restricted to spatial data, but a natural extension comes when considering spatio-temporal events, and mixtures 
between spatial events and spatio-temporal ones on another support. Mixtures of hybrids in a multivariate setting will be a welcome 
contribution. Finally, and for completeness of our proposal, adding information on covariates would enlarge the flexibility of our tools. 

\section*{Acknowledgements}
This research has been partially funded by grants UJI-B2018-04 and MTM2016-78917-R from UJI and the Spanish Ministry of Education and Science

\bibliographystyle{rss}
\bibliography{spatgraph}

\end{document}